\documentclass[twocolumn]{aastex61}

\pdfoutput=1                                    % for arXiv submission
\usepackage{amsmath,amstext}
\usepackage[T1]{fontenc}
\usepackage[figure,figure*]{hypcap}
\newcommand{\CIV}{\ion{C}{4}}
\newcommand{\HI}{\ion{H}{1}}
\newcommand{\OVI}{\ion{O}{6}}

\newcommand{\fuse}{{\sl FUSE}}
\newcommand{\kms}{\ensuremath{\mathrm{km\,s}^{-1}}}
\newcommand{\hst}{{\sl HST}}
\newcommand{\lya}{\ensuremath{\mathrm{Ly}\alpha}}
\newcommand{\Rvir}{\ensuremath{R_{\rm vir}}}

\shorttitle{Spread of Metals into the IGM}
\shortauthors{Pratt et al.}

\begin{document}

\title{The Spread of Metals into the Low-Redshift Intergalactic Medium}

\author{Cameron T. Pratt}
\affiliation{Center for Astrophysics and Space Astronomy, Department of Astrophysical and Planetary Sciences, University of Colorado, 389 UCB, Boulder, CO 80309, USA}
\author{John T. Stocke}
\affiliation{Center for Astrophysics and Space Astronomy, Department of Astrophysical and Planetary Sciences, University of Colorado, 389 UCB, Boulder, CO 80309, USA}
\author{Brian A. Keeney}
\affiliation{Center for Astrophysics and Space Astronomy, Department of Astrophysical and Planetary Sciences, University of Colorado, 389 UCB, Boulder, CO 80309, USA}
\author{Charles W. Danforth}
\affiliation{Center for Astrophysics and Space Astronomy, Department of Astrophysical and Planetary Sciences, University of Colorado, 389 UCB, Boulder, CO 80309, USA}

\begin{abstract}
We investigate the association between galaxies and metal-enriched and
metal-deficient absorbers in the local universe ($z < 0.16$) using a large
compilation of FUV spectra of bright AGN targets observed with the
Cosmic Origins Spectrograph aboard the {\sl Hubble Space Telescope}.
In this homogeneous sample of 18 \OVI\ detections at $N_{\rm O\,{\sc VI}}\geq13.5~\mathrm{cm}^{-2}$ and 
18 non-detections at $N_{\rm O\,{\sc VI}}<13.5~\mathrm{cm}^{-2}$ using \lya\ absorbers with ${N_{\rm H\,{\sc I}}\geq} 10^{14}~\mathrm{cm}^{-2}$, the maximum distance \OVI\ extends from galaxies of 
various luminosities is $\sim0.6$~Mpc, or $\sim5$ virial radii, confirming and 
refining earlier results. This is an important value that must be matched 
by numerical simulations, which input the strength of galactic winds at 
the sub-grid level. We present evidence that the primary contributors to
the spread of metals into the circum- and intergalactic media are
sub-$L^*$ galaxies ($0.25L^*<L<L^*$). The maximum distances that metals are transported
from these galaxies is comparable to, or less than, the size of a 
group of galaxies. These results suggest that, where groups are present, the metals produced by the group galaxies
do not leave the group. Since many \OVI\ non-detections in our sample occur at comparably close impact parameters 
as the metal-bearing absorbers, some more pristine intergalactic material appears to be accreting onto groups where it can mix
with metal-bearing clouds.
\end{abstract}


\section{Introduction}
\label{sec:intro}

Models of galactic evolution must incorporate the accretion of
low-metallicity gas ($Z \sim 0.1\,Z_{\Sun}$) from the ambient
intergalactic medium \citep[IGM; e.g.,][]{oppenheimer12} both in order
to resolve the ``G-dwarf problem'' \citep{pagel09} and to maintain the
high star formation rates seen in late-type galaxies like the
Milky Way \citep{binney87}. The modest metallicity of this accreting
gas suggests a source within nearby, low-mass galaxies
although the mass range of the source galaxies is not known
specifically. Ultimately, a mixture of outflows and accretion
composes the massive, gaseous halos that surround most late-type
galaxies, known as the circumgalactic medium
\citep[CGM;][]{tumlinson11,stocke13,werk14,burchett16,keeney17}.

Some studies suggest that the CGM extends from the disk of a
star-forming galaxy to its ``virial radius''
\citep[\Rvir;][]{stocke13,shull14}, which is the distance a galaxy has
gravitational influence on its surroundings. At least one
recent study presents evidence that the CGM does not extend much
beyond 1/2\,\Rvir\ \citep{prochaska17}, but \citet{shull14} argues
that parts of the CGM can include unbound outflows beyond \Rvir. 
Simulations suggest that the amount of gas and metals that escapes 
is a strong function of both galaxy mass and redshift, 
with ``gusty'' winds at high-$z$ calming down to bound 
``galactic fountains'' for the most massive halos at $z<1$ 
\citep{muratov15,muratov17,hayward17}.

The boundary between the CGM and IGM is rather ambiguous,
especially for galaxies of widely differing escape
velocities. AGN absorption-line observations support this premise
because there are no strong changes in \HI\ absorber properties,
including covering factor and mean \HI\ column density, which decline 
monotonically in the 1-5\,\Rvir\ range \citep{stocke13}. However, metal-bearing
absorbers decline rapidly away from the nearest galaxies
\citep[e.g.,][]{chen98,finn16, burchett16} and have yet to be 
detected in galaxy voids \citep{stocke07}.

Simulations by \citet{oppenheimer12} suggest that it is primarily low-mass 
galaxies whose supernova-driven winds enrich the IGM with metals, 
because winds produced by very 
massive galaxies ($M \geq 10^{11}~M_{\Sun}$) may be incapable of breaching 
their surrounding gaseous halos \citep {cote12, hayward17}. 
Instead, these outflows may fall back onto galactic 
disks and re-ignite star formation \citep*[e.g.,][]{veilleux05}, as 
in our own Galaxy \citep{keeney06,bordoloi17}. 

When large-scale simulations \citep[e.g.,][]{dave99,dave11,cen99}
place galactic winds in a cosmological context, the strength of these
winds and their full range of extent often are input at a
sub-pixel level \citep[so-called sub-grid physics, but see recent,
high-resolution simulations by the {\it FIRE} collaboration;][]{hopkins16}
so that the extent to which metals are transported away from their
source galaxy is not determined \textit{a priori} in most
simulations. Thus, this maximum extent provides both an observational bound for a galaxy's 
CGM and a constraint on galactic wind modeling within cosmological 
simulations.

In this paper, we estimate the maximum distance
winds propagate away from galaxies using low-redshift absorption
found in the far-ultraviolet (FUV) spectra of bright AGN obtained
with the {\sl Cosmic Origins Spectrograph} (COS) aboard the {\sl Hubble
Space Telescope} (\hst), in conjunction with an extensive database of
low-$z$ galaxy positions and redshifts near these sight lines
\citep[B.~Keeney et~al. 2018, in preparation]{stocke13,keeney17}. 
Since  \OVI\ (1032, 1038~\AA) exhibits the greatest extent
away from galaxies of any of the ions detected in absorption in
the FUV \citep{prochaska11,stocke13,keeney17}, this study uses only the \OVI\
doublet.

The present \OVI\ study expands upon and updates similar \OVI\ work by 
\citet{stocke06}, where a smaller sample of absorbers was used to 
determine that metals spread no more than $\sim800$~kpc from $L^*$ 
galaxies \citep[see also][]{johnson15,finn16}. For the remainder of this paper, \autoref{sec:samples} describes 
the absorption-line and galaxy survey data,
\autoref{sec:spread} presents the results, and \autoref{sec:summary}
provides a summary of our results and conclusions.

\section{Absorber and Galaxy Samples}
\label{sec:samples}

\subsection{\HI\ and \OVI\ Absorber Sample}
\label{sec:samples:absorbers}

This search for galaxy-absorber associations uses the largest survey of the low-$z$ IGM to date from \citet{danforth16}.  These authors used \hst/COS FUV spectra to construct an absorber sample along 82 AGN sight lines in the redshift range $0.05<z<0.75$. The sample includes strong \lya\ absorbers with ${N_{\rm H\,{\sc I}}\geq} 10^{14}~\mathrm{cm}^{-2}$ from \citet{danforth16}, although some \HI\ column density measurements have been revised in \citet{keeney17} and this work. This limit is high enough that Ly$\beta$ is detected for all these absorbers, increasing the accuracy of the N$_{\rm H\,{\sc I}}$ measurement. Moreover, \OVI\ (and Ly$\beta$) falls in the COS bandpass only at $z\gtrsim0.11$. Since the galaxy redshift surveys employed are naturally weighted towards $z\leq0.1$, previous data from the {\sl Far-Ultraviolet Spectroscopic Explorer} (\fuse) satellite of a few very bright AGN were incorporated to expand the range of coverage to more low-$z$ \OVI\ absorbers. The strengths of absorbers used in this study are provided in \autoref{tab:abs}.

\startlongtable
  \begin{deluxetable*}{lllccc}
  \tablecaption{Absorber Information \label{tab:abs}}
  \tablehead{\colhead{Sight Line} & \colhead{z$_{\rm abs}$} & \colhead{log N$_{\rm HI}$} & \colhead{Source} & \colhead{log N$_{\rm OVI}$} & \colhead{Source} \\
  & & \colhead{(cm$^{-2}$)} & & \colhead{(cm$^{-2}$)}}
 \startdata
 1ES 1028+511 & 0.14057 & 14.06 $\pm$ 0.18 & Danforth et al. (2016) & <13.41 & This Work \\
3C 263 & 0.06340 & 15.31 $\pm$ 0.19 & This Work & 14.52 $\pm$ 0.07 & This Work \\
3C 263 * & 0.11392  & 14.19 $\pm$ 0.12 & Danforth et al. (2016) & 13.65 $\pm$ 0.15 & Danforth et al. (2016) \\
3C 263 * & 0.12232 & 14.26 $\pm$ 0.08 & Danforth et al. (2016) & <13.23 & This Work \\
3C 263 * & 0.14075 & 14.49 $\pm$ 0.06 & Danforth et al. (2016) & 13.73 $\pm$ 0.10 & Danforth et al. (2016) \\
FBQS J1010+3003 & 0.12833 & 14.06 $\pm$ 0.32 & Danforth et al. (2016) & <13.42 & This Work \\
H 1821+643 * & 0.12120 & 14.12 $\pm$ 0.03 & Keeney et al. (2017) & <13.16 & This Work \\
HE 0226-4110 * & 0.06087 & 14.32 $\pm$ 0.10 & This Work & <13.33 & Tilton et al. (2012) \\
PG 0953+414 * & 0.06809 & 14.52 $\pm$ 0.09 & This Work & 14.35 $\pm$ 0.11 & Tilton et al. (2012) \\
PG 1001+291 & 0.11346 & 14.13 $\pm$ 0.19 & Danforth et al. (2016) & <13.33 & This Work \\
PG 1001+291 * & 0.13744 & 15.22 $\pm$ 0.30 & Danforth et al. (2016) & <13.27 & This Work \\
PG 1048+342 & 0.14471 & 14.07 $\pm$ 0.16 & Danforth et al. (2016) & <13.28 & This Work \\
PG 1116+215 & 0.13853 & 15.95 $\pm$ 0.03 & Keeney et al. (2017) & 13.78 $\pm$ 0.02 & Keeney et al. (2017) \\
PG 1216+069 A * & 0.12375 & 14.57 $\pm$ 0.05 & Keeney et al. (2017) & 14.14 $\pm$ 0.06 & Keeney et al. (2017) \\
PG 1216+069 B & 0.12375 & 14.76 $\pm$ 0.05 & Keeney et al. (2017) & 14.12 $\pm$ 0.06 & Keeney et al. (2017) \\
PG 1216+069 * & 0.12478 & 14.74 $\pm$ 0.06 & Danforth et al. (2016) & 14.17 $\pm$ 0.15 & Danforth et al. (2016) \\
PG 1216+069 * & 0.13507 & 14.75 $\pm$ 0.07 & Danforth et al. (2016) & <13.46 & This Work \\
PG 1222+216 A * & 0.15567 & 14.04 $\pm$ 0.10 & Danforth et al. (2016) & <13.40 & This Work \\
PG 1222+216 B * & 0.15567 & 14.11 $\pm$ 0.06 & Danforth et al. (2016) & <13.40 & This Work \\
PG 1259+593 & 0.00763 & 14.05 $\pm$ 0.07 & This Work & <13.30 & This Work \\
PG 1259+593 & 0.04611 & 15.45 $\pm$ 0.04 & Keeney et al. (2017) & 13.94 $\pm$ 0.12 & Keeney et al. (2017) \\
PG 1259+593 & 0.08935 & 14.11 $\pm$ 0.05 & This Work & <13.04 & This Work \\
PG 1424+240 * & 0.12134 & 15.35 $\pm$ 0.29 & Danforth et al. (2016) & 14.52 $\pm$ 0.11 & Danforth et al. (2016) \\
PG 1424+240 A * & 0.14697 & 14.60 $\pm$ 0.06 & Danforth et al. (2016) & 13.87 $\pm$ 0.23 & Danforth et al. (2016) \\
PG 1424+240 B & 0.14697 & 15.58 $\pm$ 1.41 & Danforth et al. (2016) & 13.65 $\pm$ 0.23 & Danforth et al. (2016) \\
PG 1626+554 * & 0.09382 & 14.52 $\pm$ 0.53 & This Work & <13.30 & This Work \\
PHL 1811 * & 0.07348 & 14.54 $\pm$ 0.15 & This Work & <13.03 & Tilton et al. (2012) \\
PHL 1811 * & 0.07777 & 15.40 $\pm$ 0.07 & Keeney et al. (2017) & <13.12 & Keeney et al. (2017) \\
PHL 1811 A * & 0.12060 & 14.42 $\pm$ 0.11 & This Work & <13.14 & This Work \\
PHL 1811 B * & 0.12060 & 14.33 $\pm$ 0.22 & Danforth et al. (2016) & <12.88 & This Work \\
PHL 1811 * & 0.13229 & 14.61 $\pm$ 0.01 & Keeney et al. (2017) & 13.88 $\pm$ 0.02 & Keeney et al. (2017) \\
PHL 1811 & 0.13547 & 14.98 $\pm$ 0.13 & Danforth et al. (2016) & 13.54 $\pm$ 0.16 & Danforth et al. (2016) \\
PKS 0405-123 & 0.09180 & 14.69 $\pm$ 0.03 & This Work & 13.83 $\pm$ 0.04 & Tilton et al. (2012) \\
PKS 0405-123 & 0.09655 & 14.94 $\pm$ 0.02 & This Work & 13.71 $\pm$ 0.15 & Tilton et al. (2012) \\
PKS 2005-489 & 0.01695 & 14.66 $\pm$ 0.19 & This Work & 13.76 $\pm$ 0.12 & This Work \\
PKS 2005-489 * & 0.06499 & 14.10 $\pm$ 0.22 & This Work & 13.61 $\pm$ 0.07 & Tilton et al. (2012) \\
Q 1230+0115 * & 0.07807 & 15.11 $\pm$ 0.53 & This Work & 14.47 $\pm$ 0.37 & This Work \\
SBS 1122+594 & 0.14315 & 14.33 $\pm$ 0.26 & Danforth et al. (2016) & <13.47 & This Work \\
SBS 1122+594 & 0.15545 & 15.11 $\pm$ 0.21 & Danforth et al. (2016) & 14.10 $\pm$ 0.09 & Danforth et al. (2016) \\
TON 580 & 0.13396 & 14.31 $\pm$ 0.15 & Danforth et al. (2016) & <13.47 & This Work \\
 \enddata \vspace*{0.5cm}
 \tablenotemark{A/B}{Two partially blended absorbers that are treated as a single system.}\\
 \tablenotemark{*}{Absorbers included in the ``matched'' $N_{\rm H\,{\sc I}}$ subsamples.}
\end{deluxetable*}
Only ``clean'' detections and non-detections of \OVI\ were utilized; meaning, the 
spectra exhibit no intervening lines (e.g. interstellar transitions or 
redshifted \lya\ lines) at the same wavelengths of \OVI. Also, the \OVI\ 
doublet is required to be sampled at high signal-to-noise 
($\mathrm{S/N}>15$), allowing the $\geq$ 3$\sigma$ detection of \OVI\ at 
$N_{\rm O\,{\sc VI}} \ga 10^{13.5}~\mathrm{cm}^{-2}$, or upper limits below that level. The median value for the 18 \OVI\ detections is N$_{\rm O\,{\sc VI}}=10^{13.9}~\mathrm{cm}^{-2}$ while all the 18 non-detections are at N$_{\rm O\,{\sc VI}} <$ 10$^{13.5}~\mathrm{cm}^{-2}$. 

Additionally, to permit counting only one absorber-galaxy correlation per galaxy halo, we treated any \OVI\ absorptions with $|\Delta v |<$ 250 \kms\ as a single system in the same halo \citep[see discussion of absorber systems in][]{stocke14}. While the S/N of the COS FUV spectra is 
not uniform, these high-S/N data allow a median detection of metal-enriched absorbers at the $\sim$10\% solar level; although, some metallicities in the sample may be as low as a few percent solar values based on the analysis of similar absorbers by \citet{savage14}. Starting with a \lya\ 
absorption-line redshift, there were 18 
detections and 18 non-detections in \OVI\ at $z<0.16$. 

The ``full'' samples of absorbers were created using all of the available high-S/N spectra in the \citet{danforth16} compilation. It was determined, however, that the \OVI\ non-detections have systematically smaller $N_{\rm H\,{\sc I}}$ values than the detections. In order to check for potential biases caused by this difference, we constructed ``matched'' subsamples of \OVI\ detections and non-detections by matching their $N_{\rm H\,{\sc I}}$ values within 0.2 dex for each pair. By this process, 10 matched pairs in $N_{\rm H\,{\sc I}}$ (indicated by asterisks in \autoref{tab:abs}) were created so that two subsamples could be drawn from the same parent population in $N_{\rm H\,{\sc I}}$ (Anderson-Darling $p$-value = 0.99). It was not possible to obtain a larger matched sample due to the very strong correlation between $N_{\rm H\,{\sc I}}$ and impact parameter to the nearest galaxy, which has been known since the seminal work of \citet{lanzetta95} and is almost certainly a consequence of large-scale structure formation \citep{dave99}. In \autoref{sec:spread}, results are presented using both the full samples and these samples of matched pairs.

\citet{danforth16} and \citet{keeney17} list the absorbers used in this
survey and provide spectra, velocities and equivalent widths or limits for 
\HI\ and common metal-line species, including \ion{C}{2}, \ion{C}{3},
\CIV, \ion{Si}{2}, \ion{Si}{3}, \ion{Si}{4}, and \OVI, when
they occur within the \hst/COS or \fuse\ bandpasses. Since all of
these absorbers include at least \HI\ \lya, absorber velocities
for \lya\ are used and have a velocity error of
$\pm15~\kms$ due to the absolute wavelength uncertainty
of the \hst/COS G130M and G160M gratings \citep{green12}. 
Somewhat larger velocity errors are quoted for some absorbers in 
\citet{danforth16} and \citet{keeney17} for \lya\ lines with complex line 
profiles.

\subsection{Galaxy Redshift Surveys}
\label{sec:samples:galaxies}

The galaxy data were 
obtained from four ground-based telescopes: the Sloan Digital Sky Survey 
(SDSS) spectroscopic sample \citep[DR12;][]{alam15} and
multi-object spectroscopy (MOS) from the 3.5-m Wisconsin-Indiana-Yale-NOAO 
(WIYN) telescope at Kitt Peak National Observatory, the 3.9-m 
Anglo-Australian Telescope (AAT), and the 
4-m Blanco telescope at Cerro Tololo Inter-American Observatory. While the 
SDSS spectroscopic survey is relatively shallow ($m_r\leq17.8$), it 
provides large-angle coverage which is unmatched in the south, where we 
relied upon a compilation of various galaxy redshift surveys including the 
2dF \citep{colless07} and 6dF \citep{jones09} surveys. These wide-field 
surveys were complemented by much deeper ($m_g \leq 20$) MOS, primarily 
obtained at WIYN with the HYDRA spectrograph. Individual field completeness 
levels vary but are typically $>90$\% out to completeness impact parameters ($\rho_{\rm lim}$)
of 0.5-2~Mpc for most absorbers; details of the observational process, data 
reduction and analysis, and redshift determination are presented in 
B.~Keeney et~al. (2018, in preparation). 

There are several reasons why the completeness for obtaining measurable 
	redshifts does not reach 100\%, including an inability to place fibers on 
	galaxies separated by $\leq20\arcsec$ on the sky, and very diffuse 
	galaxies whose spectrum is inconclusive despite a total 
	magnitude brighter than a given completeness limit ($L_{\rm lim}$) at the absorber redshift.

\autoref{tab:comp} presents the completeness levels of our galaxy surveys for each absorber. 
A completeness level $\geq90$\% 
	is required herein as in our first study \citet{stocke06}. Blank entries are not 
	complete to $L_{\rm lim}$ at $\geq90$\%, and so are not part of this survey. 
	Absorbers with entries of ``SDSS'' are complete to $\geq94$\% based on the limits of DR12 \citep[see][B.~Keeney et~al. 2018, in preparation]{alam15}. Some of the luminosities in \autoref{tab:ng} differ somewhat from those presented in \citet{stocke14} because they have been updated using our own photometry and analysis procedure \citep[see detailed discussion in] [] {keeney17} compared to earlier results from the literature \citep[e.g.,][]{prochaska11}.

\subsection{Nearest Galaxy Data}
	\label{sec:samples:nearest}

	Although redshift accuracies vary somewhat depending on the intrinsic galaxy 
	spectrum (e.g., pure emission-line, emission plus absorption line, or pure 
	absorption line), these errors are typically $\pm30~\kms$ as determined 
	for objects that were observed multiple times in our program (B.~Keeney et~al. 
	2018, in preparation). We \textit{de facto} assume that any galaxy within 
	$\pm1000~\kms$ of the absorber velocity could be associated with the absorber, 
	but compute a three-dimensional distance between each of these nearby galaxies 
	and the absorbers by assuming a ``reduced Hubble flow'' model. Under this assumption, the 
	line-of-sight distance between absorbers and galaxies ($D_{\rm los}$) is zero 
	where the galaxy-absorber velocity difference, $|\Delta v| \leq 400~\kms$ and is 
otherwise determined using ``pure Hubble flow''; i.e., $D_{\rm los} = (|\Delta v|-400~\kms)/H_0$. 
	
	While the ``reduced velocity'' limit is arbitrary, this choice is based upon 
	the rotation speed of an $L>L^*$ galaxy plus an additional peculiar velocity to be conservative.  Only a few galaxies with 
	$|\Delta v|>400~\kms$ are identified as nearest galaxies by this study, 
	mostly for $L>L^*$ galaxies (6~cases).
	
	We also consider scaled galaxy distances in units of \Rvir\,. With rest-frame $g$-band luminosities for all galaxies, virial radii (and halo masses) are determined from their stellar mass using a halo-matching technique described in \citet[see their Figure 1]{stocke13} and \citet{keeney17}. Figure~1 of \citet{stocke13} shows 
	the function adopted in comparison with different scaling relationships 
	used by other groups. For $L>L^*$ galaxies, these virial radii are 
	approximately a factor of two smaller than those assumed by 
	\citet{prochaska11} or the COS-Halos team 
	\citep[e.g.,][]{werk14}. Scrutiny of several dozen, low-$z$ \hst/COS-discovered absorbers finds that the identification of an absorber 
	with a specific galaxy is robust out to $\rho \la 1.4\,\Rvir$ 
	\citep{keeney17}.

We identify individual galaxies as being ``associated'' with these absorbers, but it is possible that some absorbers are actually affiliated with entire groups of galaxies in which the nearest galaxy is a member of the group \citep[discussed in detail in][]{stocke14}. Since virtually {\bf all} galaxies are in groups of some size (see Local Supercluster studies by \citealt{tully09}), it is difficult to determine whether the absorber is most closely associated with an individual galaxy, particularly when $\rho \geq 1.4$ \Rvir\ (\citealt{keeney17}; see \citealt{stocke17} for discussions of a specific case study). Moreover, there is no statistically meaningful way to discriminate between absorbers associated with groups vs. individual galaxies or to know which halo mass distribution these absorbers should be connected with.  This is an ambiguity for all studies concerning absorber-galaxy connections at both low- and high-$z$ \citep[e.g.,][]{steidel10, werk14}. Regardless, this study utilizes individual galaxy virial radii to estimate how far metals are spread from their putative source galaxy.

	\autoref{tab:ng} presents the basic data used in this study, in which the third column 
	identifies the \OVI\ detections and non-detections. The remaining columns 
	(keyed to the limiting galaxy luminosity, $L_{\rm lim}$) list the nearest 
	galaxy luminosity, $L$, and three-dimensional absorber-galaxy physical distance, $D$, and
distances scaled by \Rvir\,. 
	Here we identify the nearest galaxy in two different ways: using the smallest physical distance and the smallest distance scaled by \Rvir\,. Conceivably, it is possible for a very luminous galaxy to be the nearest galaxy to an absorber in terms of \Rvir\ even if it is physically farther away than a lower-luminosity galaxy. Columns~4-7
	use all available data regardless of galaxy luminosity. Columns~8-11, 12-15, and 16-19 use all 
	galaxies with $L\geq 0.25\,L^*$, 
	$L\geq 0.5\,L^*$, and $L\geq L^*$ respectively. 
	Absorber regions that were not surveyed deeply enough to reach the limiting 
	luminosities given at the top have no data shown.

%\clearpage
%\afterpage{
\startlongtable
  \begin{deluxetable*}{lcccccccccc}
    \tablecaption{Galaxy Survey Completeness Limits Surrounding IGM Absorbers \label{tab:comp}}
    \tablehead{\colhead{} & \colhead{} & \colhead{}&  
	       \multicolumn{2}{c}{$L_{\rm lim} = 0.25\,L^{*}$} && \multicolumn{2}{c}{$L_{\rm lim} = 0.5\,L^{*}$} &&
               \multicolumn{2}{c}{$L_{\rm lim} = L^{*}$} \\
	       \cline{4-5} \cline{7-8} \cline{10-11} 
	       \colhead{Sight Line} & \colhead{$z_{\rm abs}$} &&
	       \colhead{$\rho_{\rm lim}$} & \colhead{Completeness} &&
	       \colhead{$\rho_{\rm lim}$} & \colhead{Completeness} &&
	       \colhead{$\rho_{\rm lim}$} & \colhead{Completeness} \\
               & & &
               \colhead{(Mpc)} & \colhead{(\%)} &&
	       \colhead{(Mpc)} & \colhead{(\%)} &&
	       \colhead{(Mpc)} & \colhead{(\%)}}
\startdata
1ES 1028+511 & 0.14057 && \nodata & \nodata &&  3.00   &  100\%  &&  3.00   &  100\%  \\
3C 263 & 0.06340 &&  1.48   &  100\%  &&  SDSS   & $\rm SDSS$ &&  SDSS   & $\rm SDSS$ \\
3C 263 & 0.11392 &&  2.51   & 92.65\% &&  2.51   &  100\%  &&  SDSS   & $\rm SDSS$ \\
3C 263 & 0.12232 &&  2.67   & 94.38\% &&  2.67   &  100\%  &&  SDSS   & $\rm SDSS$ \\
3C 263 & 0.14075 && \nodata & \nodata &&  3.01   &  100\%  &&  3.01   &  100\%  \\
FBQS J1010+3003 & 0.12833 &&  2.78   & 95.73\% &&  2.78   &  100\%  &&  2.78   &  100\%  \\
H 1821+643 & 0.12120 &&  2.65   & 93.07\% &&  2.65   & 96.15\% &&  SDSS   & $\rm SDSS$ \\
HE 0226-4110 & 0.06087 &&  1.42   &  100\%  &&  1.42   &  100\%  &&  1.42   &  100\%  \\
PG 0953+414 & 0.06809 &&  1.58   &  100\%  &&  SDSS   & $\rm SDSS$ &&  SDSS   & $\rm SDSS$ \\
PG 1001+291 & 0.11346 &&  2.50   & 94.87\% &&  2.50   & 96.15\% &&  SDSS   & $\rm SDSS$ \\
PG 1001+291 & 0.13744 && \nodata & \nodata &&  2.95   & 95.92\% && \nodata & \nodata \\
PG 1048+342 & 0.14471 && \nodata & \nodata &&  3.08   &  95\%   &&  2.00   &  100\%  \\
PG 1116+215 & 0.13853 && \nodata & \nodata && \nodata & \nodata && \nodata & \nodata \\
PG 1216+069 & 0.12375 &&  1.35   & 95.24\% &&  2.16   & 92.86\% &&  2.16   &  100\%  \\
PG 1216+069 & 0.12478 &&  1.36   & 95.45\% &&  2.17   & 93.33\% &&  2.17   &  100\%  \\
PG 1216+069 & 0.13507 && \nodata & \nodata &&  1.89   & 90.91\% &&  2.32   &  100\%  \\
PG 1222+216 & 0.15567 && \nodata & \nodata && \nodata & \nodata &&  2.78   &  100\%  \\
PG 1259+593 & 0.00763 &&  SDSS   & $\rm SDSS$ &&  SDSS   & $\rm SDSS$ &&  SDSS   & $\rm SDSS$ \\
PG 1259+593 & 0.04611 &&  SDSS   & $\rm SDSS$ &&  SDSS   & $\rm SDSS$ &&  SDSS   & $\rm SDSS$ \\
PG 1259+593 & 0.08935 &&  2.02   & 91.43\% && \nodata & \nodata &&  SDSS   & $\rm SDSS$ \\
PG 1424+240 & 0.12134 && \nodata & \nodata &&  2.65   & 95.45\% &&  SDSS   & $\rm SDSS$ \\
PG 1424+240 & 0.14697 && \nodata & \nodata &&  2.65   & 93.55\% &&  3.12   &  100\%  \\
PG 1626+554 & 0.09382 &&  0.84   & 90.91\% &&  0.84   &  100\%  &&  SDSS   & $\rm SDSS$ \\
PHL 1811 & 0.07348 &&  1.70   & 97.78\% &&  1.70   &  100\%  &&  1.70   &  100\%  \\
PHL 1811 & 0.07777 &&  1.78   & 98.11\% &&  1.78   &  100\%  &&  1.78   &  100\%  \\
PHL 1811 & 0.12060 &&  1.05   & 92.31\% &&  2.64   & 98.65\% &&  2.64   &  100\%  \\
PHL 1811 & 0.13229 &&  1.14   & 92.59\% &&  2.85   & 96.84\% &&  2.85   &  100\%  \\
PHL 1811 & 0.13547 &&  1.17   & 92.59\% &&  2.91   & 96.94\% &&  2.91   &  100\%  \\
PKS 0405-123 & 0.09180 &&  1.87   &  100\%  &&  1.87   &  100\%  &&  1.87   &  100\%  \\
PKS 0405-123 & 0.09655 &&  1.95   & 91.67\% &&  1.95   &  100\%  &&  1.95   &  100\%  \\
PKS 2005-489 & 0.01695 && \nodata & \nodata && \nodata & \nodata && \nodata & \nodata \\
PKS 2005-489 & 0.06499 &&  0.15   &  100\%  &&  0.15   &  100\%  &&  0.15   &  100\%  \\
Q 1230+0115 & 0.07807 &&  1.79   &  100\%  &&  SDSS   & $\rm SDSS$ &&  SDSS   & $\rm SDSS$ \\
SBS 1122+594 & 0.14315 && \nodata & \nodata &&  3.05   & 93.65\% &&  1.52   &  100\%  \\
SBS 1122+594 & 0.15545 && \nodata & \nodata &&  3.27   & 93.75\% &&  2.94   & 92.31\% \\
TON 580 & 0.13396 &&  2.88   &  100\%  &&  2.88   &  100\%  &&  2.88   &  100\%  \\
\enddata
\end{deluxetable*}

%\afterpage{
\begin{longrotatetable}
  \begin{deluxetable*}{lccccccccccccccccccccc}
    \tablecaption{Distances from Absorbers to their Nearest Galaxies \label{tab:ng}}
    \tablehead{ 
      & & & 
      \multicolumn{4}{c}{No Luminosity Limit}       && \multicolumn{4}{c}{$L_{\rm lim} = 0.25\,L^*$} &&
      \multicolumn{4}{c}{$L_{\rm lim} = 0.5\, L^*$} && \multicolumn{4}{c}{$L_{\rm lim} = L^*$} \\
      \cline{4-7}  \cline{9-12} \cline{14-17} \cline{19-22} & & & 
      \multicolumn{2}{c}{Physical} & \multicolumn{2}{c}{Scaled by \Rvir} &&
      \multicolumn{2}{c}{Physical} & \multicolumn{2}{c}{Scaled by \Rvir} &&
      \multicolumn{2}{c}{Physical} & \multicolumn{2}{c}{Scaled by \Rvir} &&
      \multicolumn{2}{c}{Physical} & \multicolumn{2}{c}{Scaled by \Rvir} \\
      \colhead{Sight Line} & \colhead{$\rm  z_{abs}$} & \colhead{\OVI} 
      & 
      \colhead{$L$} & \colhead{$D$} & \colhead{$L$} & \colhead{$D/\Rvir$}
      && 
      \colhead{$L$} & \colhead{$D$} & \colhead{$L$} & \colhead{$D/\Rvir$}
      &&  
      \colhead{$L$} & \colhead{$D$} & \colhead{$L$} & \colhead{$D/\Rvir$}
      &&
      \colhead{$L$} & \colhead{$D$} & \colhead{$L$} & \colhead{$D/\Rvir$}}
    \startdata
    1ES 1028+511 & 0.14057 & n &  0.328  &  1.53   &  1.94   &  9.65   & & \ldots  & \ldots  & \ldots  & \ldots  & &  1.94   &  2.19   &  1.94   &  9.65   & &  1.94   &  2.19   &  1.94   &  9.65   \\
3C 263 & 0.06340 & y &  0.283  &  0.064  &  0.283  &  0.538  & &  0.283  &  0.064  &  0.283  &  0.538  & &  0.922  &  1.53   &  0.922  &  8.70   & &  2.95   &  4.06   &  2.95   &  15.6   \\
3C 263 & 0.11392 & y &  0.368  &  0.354  &  0.368  &  2.70   & &  0.368  &  0.354  &  0.368  &  2.70   & &  0.588  &  0.709  &  0.588  &  4.66   & &  1.15   &  0.987  &  1.15   &  5.19   \\
3C 263 & 0.12232 & n &  0.430  &  1.11   &  0.430  &  8.11   & &  0.430  &  1.11   &  0.430  &  8.11   & & \ldots  & \ldots  & \ldots  & \ldots  & &  1.39   &  3.98   &  1.39   &  19.7   \\
3C 263 & 0.14075 & y &  1.87   &  0.622  &  1.87   &  2.79   & & \ldots  & \ldots  & \ldots  & \ldots  & &  1.87   &  0.622  &  1.87   &  2.79   & &  1.87   &  0.622  &  1.87   &  2.79   \\
FBQS J1010+3003 & 0.12833 & n &  0.235  &  0.529  &  1.49   &  3.18   & &  1.49   &  0.658  &  1.49   &  3.18   & &  1.49   &  0.658  &  1.49   &  3.18   & &  1.49   &  0.658  &  1.49   &  3.18   \\
H 1821+643 & 0.12120 & n &  0.776  &  0.159  &  0.776  &  0.952  & &  0.776  &  0.159  &  0.776  &  0.952  & &  0.776  &  0.159  &  0.776  &  0.952  & &  1.23   &  1.16   &  1.23   &  5.95   \\
HE 0226-4110 & 0.06087 & n &  0.287  &  0.352  &  0.287  &  2.93   & &  0.287  &  0.352  &  0.287  &  2.93   & &  0.951  &  1.29   &  0.951  &  7.20   & & \ldots  & \ldots  & \ldots  & \ldots  \\
PG 0953+414 & 0.06809 & y &  0.892  &  0.611  &  0.892  &  3.49   & &  0.892  &  0.611  &  0.892  &  3.49   & &  0.892  &  0.611  &  0.892  &  3.49   & &  1.04   &  1.07   &  1.04   &  5.81   \\
PG 1001+291 & 0.11346 & n &  0.225  &  0.816  &  1.01   &  6.08   & &  0.324  &  0.883  &  1.01   &  6.08   & &  1.01   &  1.11   &  1.01   &  6.08   & &  1.01   &  1.11   &  1.01   &  6.08   \\
PG 1001+291 & 0.13744 & n &  0.132  &  0.055  &  0.132  &  0.585  & & \ldots  & \ldots  & \ldots  & \ldots  & &  1.78   &  0.704  &  1.78   &  3.20   & & \ldots  & \ldots  & \ldots  & \ldots  \\
PG 1048+342 & 0.14471 & n &  0.364  &  0.410  &  0.364  &  3.15   & & \ldots  & \ldots  & \ldots  & \ldots  & &  0.935  &  1.01   &  2.68   &  5.16   & &  2.68   &  1.31   &  2.68   &  5.16   \\
PG 1116+215 & 0.13853 & y &  1.76   &  0.139  &  1.76   &  0.632  & & \ldots  & \ldots  & \ldots  & \ldots  & & \ldots  & \ldots  & \ldots  & \ldots  & & \ldots  & \ldots  & \ldots  & \ldots  \\
PG 1216+069 & 0.12375 & y &  0.657  &  0.094  &  0.657  &  0.595  & &  0.657  &  0.094  &  0.657  &  0.595  & &  0.657  &  0.094  &  0.657  &  0.595  & &  1.45   &  0.701  &  1.45   &  3.40   \\
PG 1216+069 & 0.12478 & y &  0.657  &  0.094  &  0.657  &  0.595  & &  0.657  &  0.094  &  0.657  &  0.595  & &  0.657  &  0.094  &  0.657  &  0.595  & &  1.45   &  0.706  &  1.45   &  3.43   \\
PG 1216+069 & 0.13507 & n &  1.38   &  0.758  &  1.38   &  3.75   & & \ldots  & \ldots  & \ldots  & \ldots  & &  1.38   &  0.758  &  1.38   &  3.75   & &  1.38   &  0.758  &  1.38   &  3.75   \\
PG 1222+216 & 0.15567 & n &  0.646  &  0.498  &  0.646  &  3.17   & & \ldots  & \ldots  & \ldots  & \ldots  & & \ldots  & \ldots  & \ldots  & \ldots  & &  2.04   &  1.89   &  2.04   &  8.19   \\
PG 1259+593 & 0.00763 & n &  0.085  &  0.474  &  0.770  &  3.51   & &  0.770  &  0.586  &  0.770  &  3.51   & &  0.770  &  0.586  &  0.770  &  3.51   & &  1.02   &  3.40   &  4.24   &  13.0   \\
PG 1259+593 & 0.04611 & y &  0.065  &  0.089  &  0.506  &  0.958  & &  0.506  &  0.138  &  0.506  &  0.958  & &  0.506  &  0.138  &  0.506  &  0.958  & &  2.37   &  0.326  &  2.37   &  1.35   \\
PG 1259+593 & 0.08935 & n &  0.198  &  1.76   &  0.429  &  12.9   & &  0.429  &  1.76   &  0.429  &  12.9   & & \ldots  & \ldots  & \ldots  & \ldots  & &  1.69   &  4.65   &  1.69   &  21.4   \\
PG 1424+240 & 0.12134 & y &  1.06   &  0.201  &  1.06   &  1.09   & & \ldots  & \ldots  & \ldots  & \ldots  & &  1.06   &  0.201  &  1.06   &  1.09   & &  1.06   &  0.201  &  1.06   &  1.09   \\
PG 1424+240 & 0.14697 & y &  0.916  &  0.497  &  0.916  &  2.82   & & \ldots  & \ldots  & \ldots  & \ldots  & &  0.916  &  0.497  &  0.916  &  2.82   & &  2.28   &  1.60   &  2.28   &  6.69   \\
PG 1626+554 & 0.09382 & n &  1.06   &  2.55   &  1.06   &  13.7   & & \ldots  & \ldots  & \ldots  & \ldots  & & \ldots  & \ldots  & \ldots  & \ldots  & &  1.06   &  2.55   &  1.06   &  13.7   \\
PHL 1811 & 0.07348 & n &  0.131  &  0.344  &  2.97   &  1.95   & &  0.620  &  0.446  &  2.97   &  1.95   & &  0.620  &  0.446  &  2.97   &  1.95   & &  2.97   &  0.508  &  2.97   &  1.95   \\
PHL 1811 & 0.07777 & n &  0.076  &  0.235  &  0.246  &  2.72   & &  0.695  &  0.542  &  3.23   &  3.01   & &  0.695  &  0.542  &  3.23   &  3.01   & &  1.29   &  0.790  &  3.23   &  3.01   \\
PHL 1811 & 0.12060 & n &  0.169  &  1.23   &  0.169  &  12.2   & & \ldots  & \ldots  & \ldots  & \ldots  & & \ldots  & \ldots  & \ldots  & \ldots  & & \ldots  & \ldots  & \ldots  & \ldots  \\
PHL 1811 & 0.13229 & y &  0.645  &  0.228  &  0.645  &  1.45   & &  0.645  &  0.228  &  0.645  &  1.45   & &  0.645  &  0.228  &  0.645  &  1.45   & &  2.65   &  2.39   &  2.65   &  9.53   \\
PHL 1811 & 0.13547 & y &  0.150  &  0.519  &  0.150  &  5.30   & &  0.333  &  1.02   &  1.14   &  6.04   & &  0.771  &  1.07   &  1.14   &  6.04   & &  1.14   &  1.15   &  1.14   &  6.04   \\
PKS 0405-123 & 0.09180 & y &  0.010  &  0.135  &  0.010  &  2.37   & &  0.345  &  0.449  &  0.345  &  3.51   & & \ldots  & \ldots  & \ldots  & \ldots  & & \ldots  & \ldots  & \ldots  & \ldots  \\
PKS 0405-123 & 0.09655 & y &  0.739  &  0.270  &  0.739  &  1.65   & &  0.739  &  0.270  &  0.739  &  1.65   & &  0.739  &  0.270  &  0.739  &  1.65   & &  1.93   &  0.934  &  1.93   &  4.15   \\
PKS 2005-489 & 0.01695 & y &  0.002  &  0.255  &  1.15   &  3.30   & & \ldots  & \ldots  & \ldots  & \ldots  & & \ldots  & \ldots  & \ldots  & \ldots  & & \ldots  & \ldots  & \ldots  & \ldots  \\
PKS 2005-489 & 0.06499 & y &  0.024  &  0.529  &  15.2   &  5.52   & & \ldots  & \ldots  & \ldots  & \ldots  & & \ldots  & \ldots  & \ldots  & \ldots  & & \ldots  & \ldots  & \ldots  & \ldots  \\
Q 1230+0115 & 0.07807 & y &  0.175  &  0.054  &  0.175  &  0.529  & &  0.652  &  0.528  &  0.652  &  3.36   & &  0.652  &  0.528  &  0.652  &  3.36   & &  1.12   &  1.52   &  2.40   &  7.92   \\
SBS 1122+594 & 0.14315 & n &  0.599  &  0.529  &  0.599  &  3.46   & & \ldots  & \ldots  & \ldots  & \ldots  & &  0.599  &  0.529  &  0.599  &  3.46   & & \ldots  & \ldots  & \ldots  & \ldots  \\
SBS 1122+594 & 0.15545 & y &  0.577  &  0.115  &  0.577  &  0.762  & & \ldots  & \ldots  & \ldots  & \ldots  & &  0.577  &  0.115  &  0.577  &  0.762  & &  2.53   &  1.74   &  2.53   &  7.03   \\
TON 580 & 0.13396 & n &  0.456  &  0.859  &  0.456  &  6.14   & &  0.456  &  0.859  &  0.456  &  6.14   & &  1.05   &  1.67   &  2.16   &  7.80   & &  1.05   &  1.67   &  2.16   &  7.80   \\
    \enddata
    \vspace*{0.3cm}
  \end{deluxetable*}
\end{longrotatetable}

\section{The Spread of \OVI\ around Low-$z$ Galaxies} 
\label{sec:spread}

\begin{figure}
\includegraphics[width=0.5\textwidth]{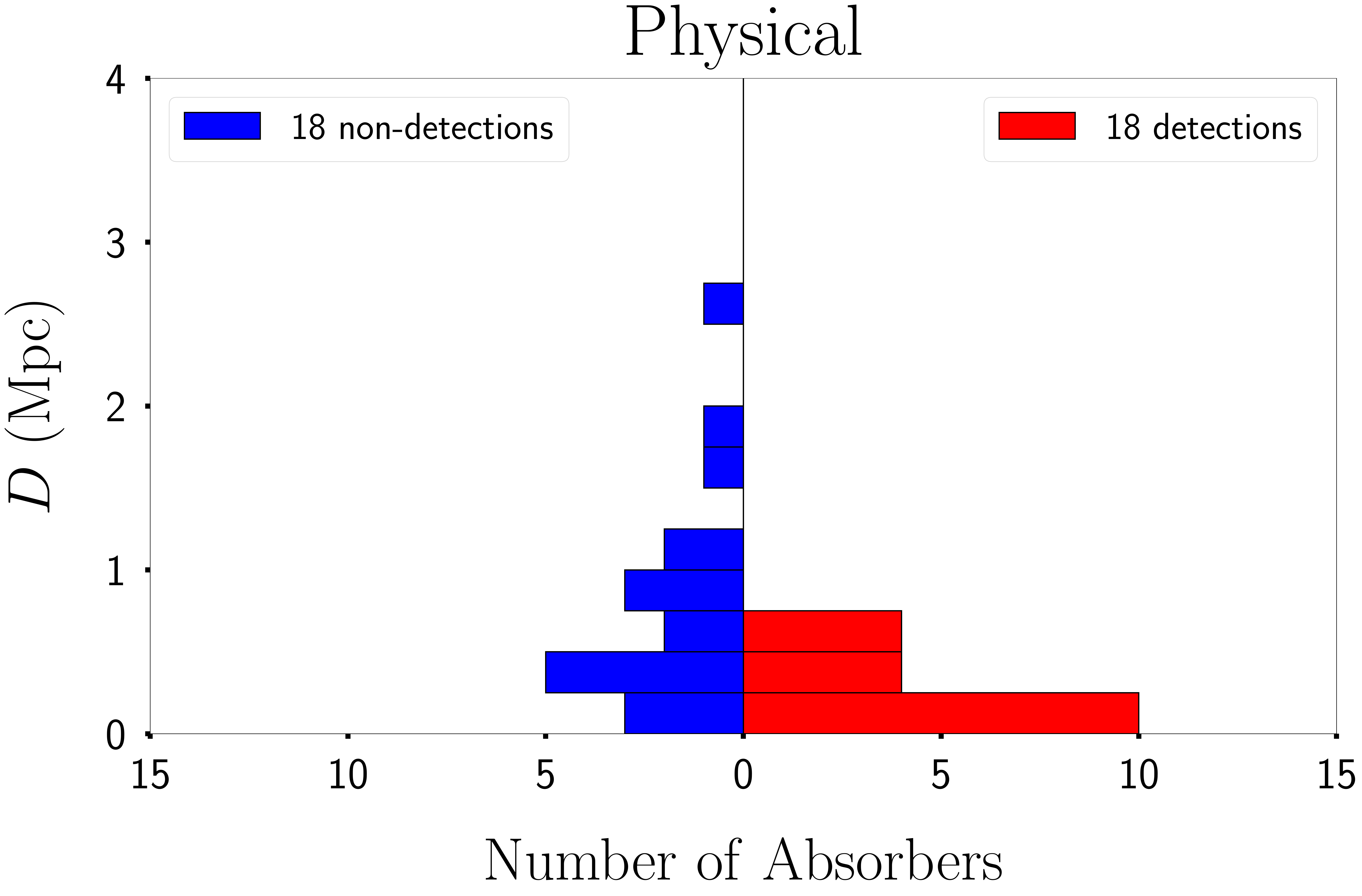}
\par \vspace{1em}
\includegraphics[width=0.5\textwidth]{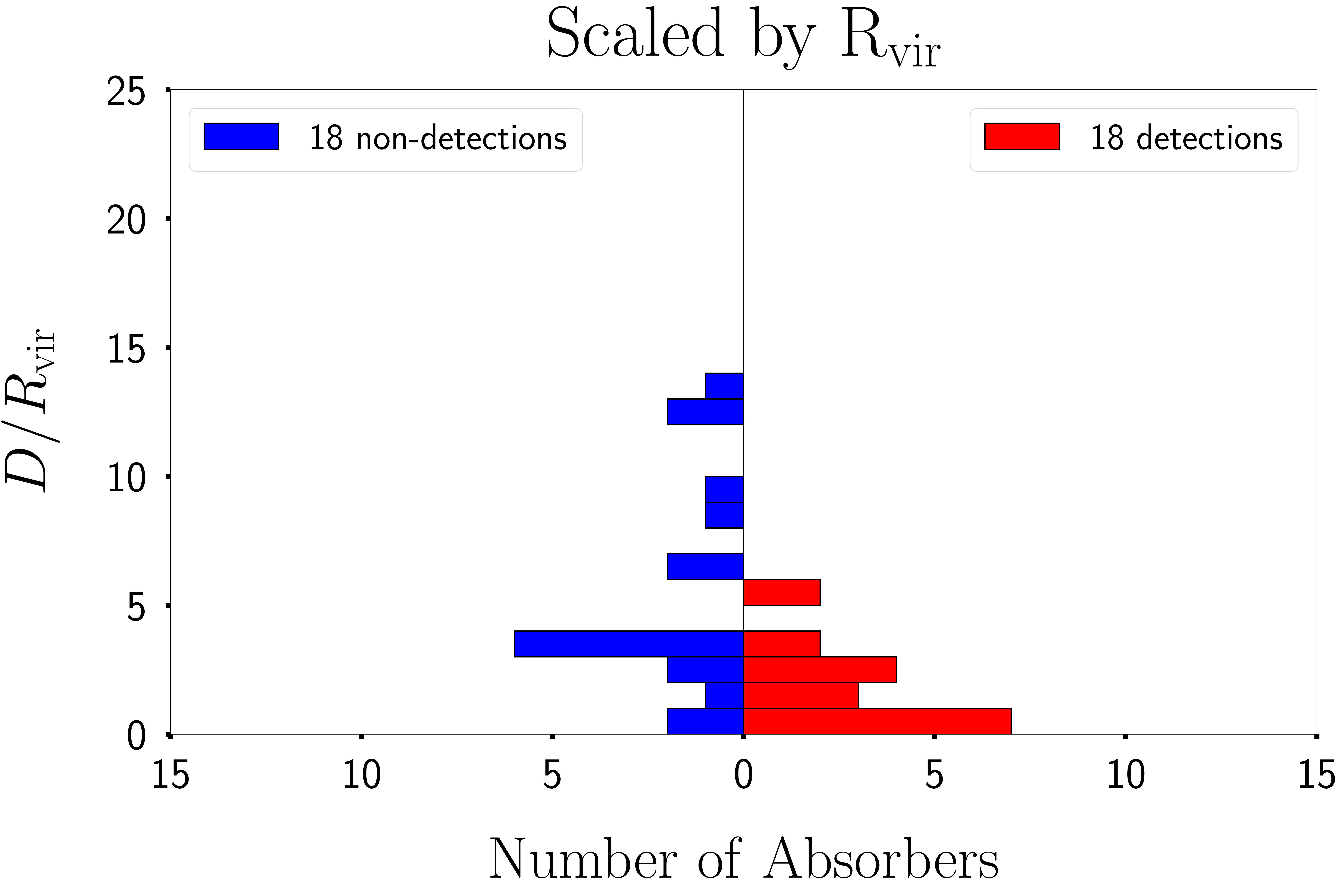}
\vspace{-2em}
\caption{Histograms of \OVI\ nearest galaxy distances in Mpc (\textit{top}) and scaled nearest galaxy distances in units of $R_{\rm vir}$ (\textit{bottom}). The red bars to the right represent the \OVI\ detections, and the blue bars to the left are non-detections. No galaxy luminosity limit was applied for these data, thus utilizing all of the galaxy redshift data we have available (i.e., columns~4-7 of \autoref{tab:ng}).}
\label{fig:OVI_hist}
\end{figure}

The most straightforward way to  examine the spread of \OVI\ from galaxies 
is through \autoref{fig:OVI_hist} which shows the 18 \OVI\ 
detections and 18 \OVI\ non-detections in double-sided histograms. The top histogram 
shows that \OVI\ spreads no more than the 750~kpc 
bin from the nearest galaxy; the specific largest value is 620~kpc. The median luminosity of the galaxy 
physically nearest to metal-enriched IGM absorbers is $0.61\,L^*$; this 
median becomes $0.66\,L^*$ when considering galaxies nearest to absorbers in units of \Rvir\,. 
Due to the large variation in nearest galaxy luminosities and thus virial radii, 
the metal spread extends to as much as 5\,\Rvir\ in some cases, leaving little 
doubt that the most remote absorbers are unbound gravitationally from the 
nearest galaxy.

% Figures 1 & 2

\begin{figure*}
\includegraphics[width=0.4\textwidth]{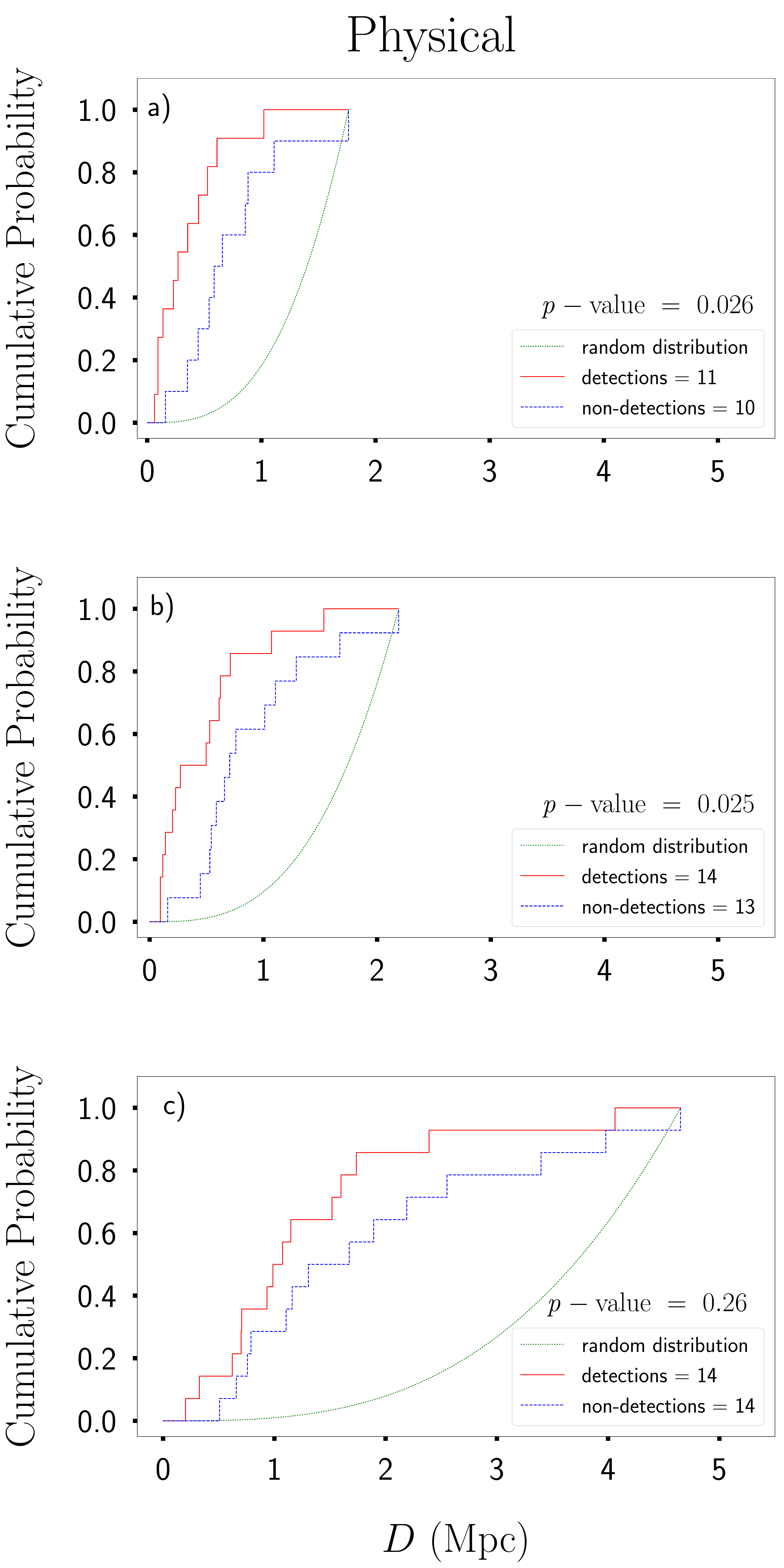}
\includegraphics[width=0.2\textwidth]{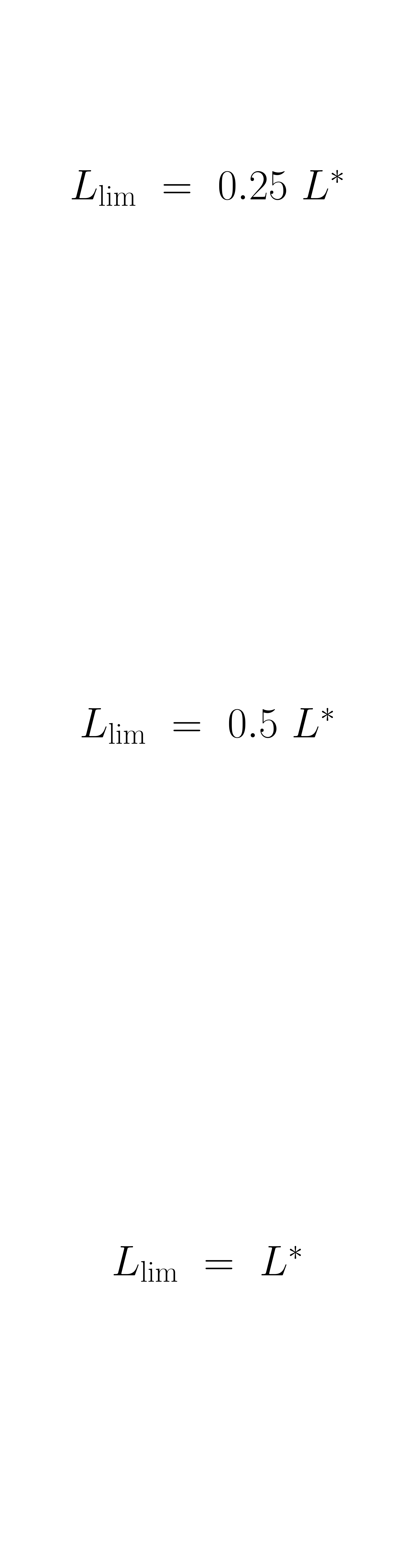}
\includegraphics[width=0.4\textwidth]{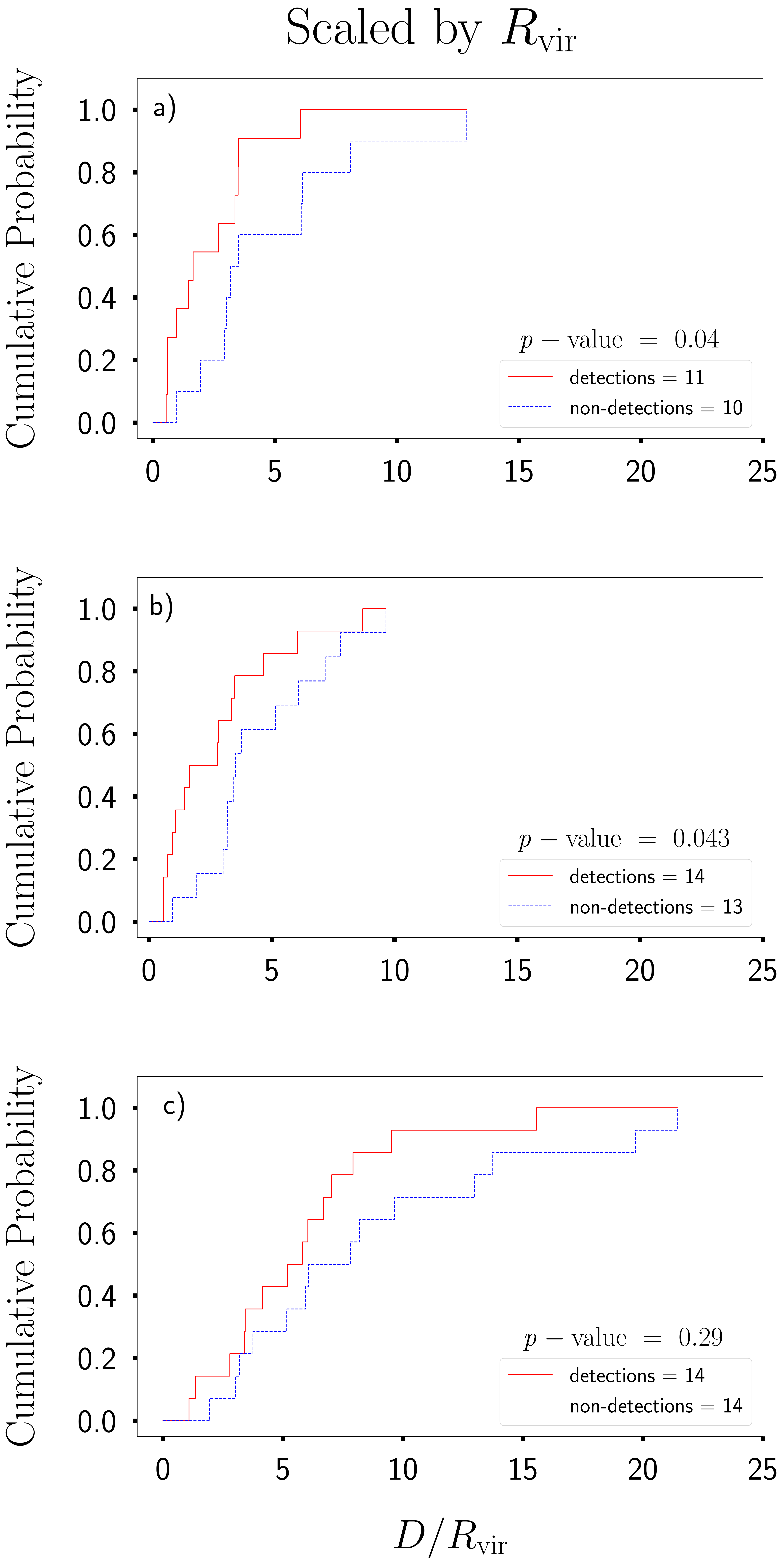}
\vspace{-2em}
\caption{Cumulative distribution functions of \OVI\ nearest galaxy distances (\textit{left}) and distances scaled by \Rvir\ (\textit{right}). The solid red lines represent \OVI\ detections, the dashed blue lines are \OVI\ non-detections, and the dotted green lines represent a random distribution of galaxies within a similar volume. The different panels (\textit{a-c}) show the limiting luminosities of 0.25, 0.5, and $1\,L^*$. All panels show a {\it{p}}-value found by the Anderson-Darling test, used to compare the distributions of \ion{O}{6} detections to non-detections given the full samples of absorbers.}
\label{fig:OVI_edf}
\end{figure*}

Using the ``matched'' subsamples of 10 \OVI\ detections and 10 non-detections, the impact parameter distribution shown in \autoref{fig:OVI_hist} remains largely unchanged; i.e., the \OVI\ absorbers are closer to their associated galaxy, but the statistical difference is smaller ($p$-value = 0.27). The lower probability ($\approx 70\%$) of being drawn from different parent populations is due primarily to the much smaller sample size of ten pairs only. This statement was verified through the fiat of creating new subsamples by counting each detection and non-detection twice. This reduced the Anderson-Darling $p$-value from 0.27 to 0.05. 
Similar to the full samples, however, the matched samples contain no \OVI\ absorbers beyond $\rho=0.6$ Mpc while nearly 50\% of the \OVI\ non-detections are at $\rho >$ 0.6 Mpc, including one \OVI\,-deficient absorber at $\rho >$ 2 Mpc.

To determine the maximum spread of metals more robustly, we show cumulative distribution functions (CDFs) in \autoref{fig:OVI_edf} for three different 
$L_{\rm lim}$ cuts in our galaxy survey where the sampling is 
complete to $\geq 90$\%. Absorbers are included in each 
luminosity bin \textbf{only if} the galaxy survey is complete at or below $L_{\rm lim}$.

The Anderson-Darling test was then used to determine the likelihood these two distributions were randomly drawn from the same parent population. 
For the $L\geq L^*$ subsample, there is virtually no difference between the 
distributions of galaxies nearest to \OVI\ detections and non-detections. The 
greatest contrast ($p$-value = 0.025) between the detections and non-detections 
is for the $L\geq0.5\,L^*$ subsample, for which the median distances to \OVI\ 
detections (0.38~Mpc and 2.22 \Rvir) are considerably smaller than those for non-detections (0.70~Mpc and 3.51 \Rvir). 

When applying these same galaxy survey constraints to the N$_{\rm H\,{\sc I}}$ matched samples of absorbers, the sizes of \OVI\ detections and non-detections decrease to single digits for each luminosity cut. Similar to the full samples, the impact parameters to \OVI\ detections are systematically smaller than non-detections. The reduced sample sizes, however, render any statistical tests for differences quite uncertain; e.g., the $L>0.5L^*$ samples differ only with a $p$-value = 0.24. While the results from the matched samples support the inference that \OVI\ detections are more closely associated with galaxies than non-detections, they do not do so strongly. Therefore, we now turn our attention to those tests which use only the full sample of \OVI\ detections.

By using only those 11~absorbers in the full sample of \OVI\ detections, whose surroundings were surveyed to at least 
$0.25\,L^*$, we inspected the luminosities of the physically nearest galaxies: there are four $L< 0.5\,L^*$ galaxies, seven $0.5\,L^*\leq L<L^*$ galaxies, and zero $L\geq L^*$ galaxies. This is 
intriguing evidence that the primary contributor to 
spreading metals into the CGM/IGM are galaxies somewhat fainter than $L^*$, similar 
to the Milky Way and M33.

To better compare our results for different  $L_{\rm lim}$, we scale the absorber-galaxy distance by the mean distance 
between galaxies of equal or greater luminosity ($\langle D_{\rm int} \rangle$), given by the inverse cube root of the integral galaxy 
luminosity function. The SDSS luminosity function of 
\citet{montero-dorta09} with $K$-corrections from \citet*{chilingarian10} and 
\citet{chilingarian12} were used to find values of:
$\langle D_{\rm int} \rangle = 5.96$, 7.45, and 10.4~Mpc for 
$L_{\rm lim} = 0.25$, 0.5, and $1\,L^*$, respectively.

\autoref{fig:OVI_lf} compares the CDFs of the \OVI\ detections for the three 
different $L_{\rm lim}$ subsamples. The median value of 
$D/\langle D_{\rm int}\rangle$ for the $L_{\rm lim}=0.5L^*$ subsample finds that these galaxies are 
$\sim20$ times closer to absorbers than they are to other 
$L\geq0.5\,L^*$ galaxies. Clearly, \OVI\ absorbers are tightly correlated 
with sub-$L^*$ galaxies.

\section{Summary of Results and Discussion} 
\label{sec:summary}

\begin{figure}
\includegraphics[width=\columnwidth]{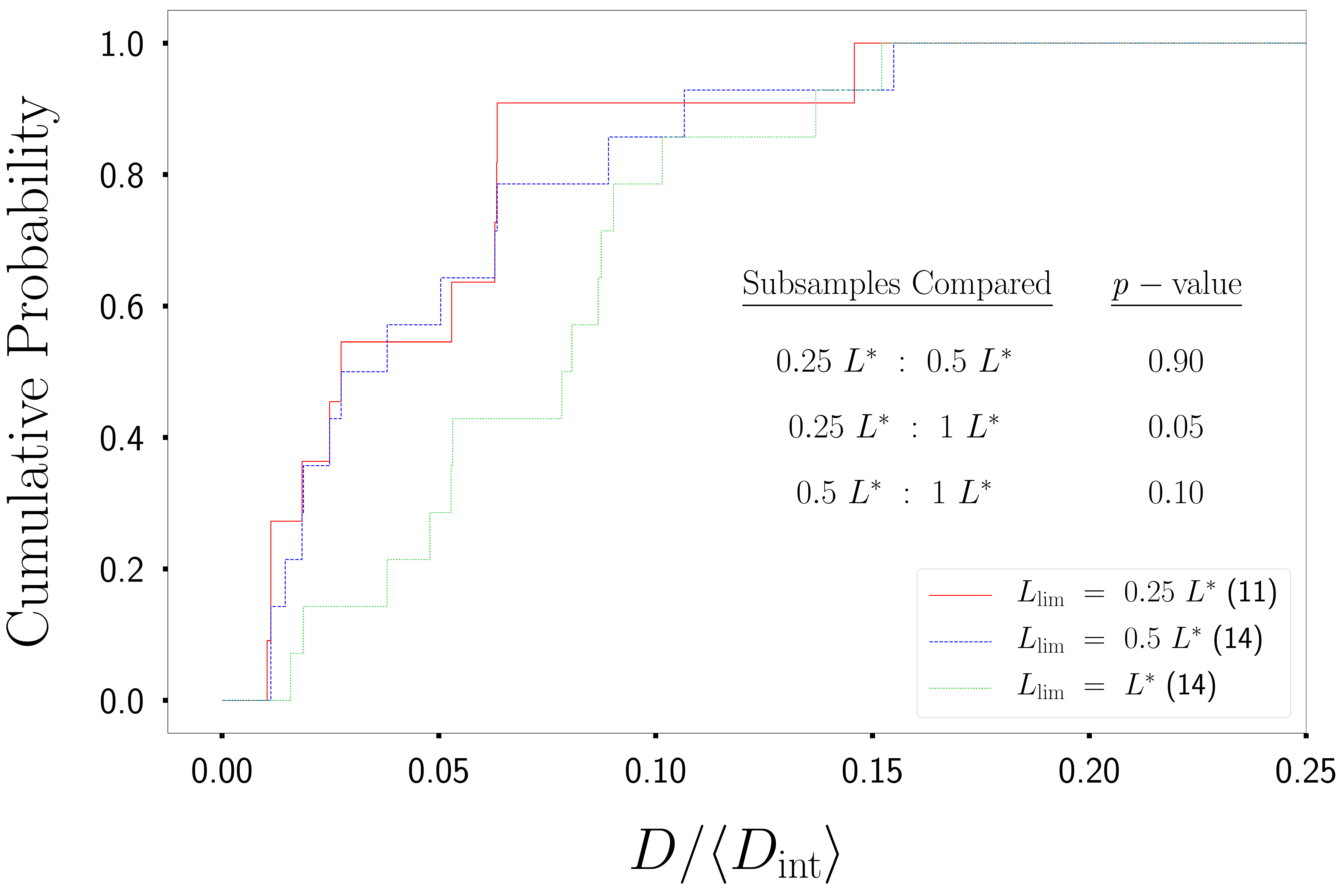}
\vspace{-2em}
\caption{Cumulative distribution functions of \OVI\ nearest galaxy distances scaled by the mean distances between galaxies of similar or greater luminosity. The red, blue, and green lines show the results for $L_{\rm lim} = 0.25$, 0.5, and $1\,L^*$, respectively. The Anderson-Darling $p$-values comparing the subsamples are shown in the figure.}
\label{fig:OVI_lf}
\end{figure}

Based on high-S/N, high-resolution FUV spectroscopy for samples of low-$z$ \OVI\ absorption-line detections and 
non-detections, we find that \OVI\ is not detected 
beyond a physical distance of $\sim0.6$~Mpc, or $\sim5\,\Rvir$ from the 
nearest galaxy. These results are in good agreement with those found by 
\citet{stocke06}, who reported a spread of \OVI\ to a maximum physical 
distance of $\sim800$~kpc, or 3.5-5\,\Rvir\ from $L\geq L^*$ galaxies. 
More recently, \citet*{johnson15} reported \OVI\ detections at 1-3\,\Rvir\ 
around galaxies of luminosities $L>0.1\,L^*$ at $z<0.4$. Also, the correlation 
lengths found here are similar to those found by \citet{finn16} in a 
large, statistical study of low-$z$ \OVI\ absorbers. Since these other recent 
studies use \hst/{\sl{COS}} FUV spectra of comparable or lesser S/N as the present 
study, these conclusions are all limited to modest metallicity gas \citep [a median level of $Z\sim0.1Z_{\odot}$ is suggested for this sample by the results of] [] {savage14}; lower metallicity gas may be more pervasive in the 
IGM at both low- and high-$z$ \citep*[e.g.,][]{aguirre02} than as measured here. Despite significant differences between the CDFs of \OVI\ detections and non-detections, many \OVI\ non-detections 
are found at comparable impact parameters to the detections as shown in \autoref{fig:OVI_edf} and \autoref{fig:OVI_lf}. We interpret this result as an indication of rather pristine gas falling onto galaxies from the IGM \citep[see also][]{stocke13}. 

By creating subsamples, defined by $L_{\rm lim}$, we find evidence that \OVI\ absorbers are more tightly correlated with $L<L^*$ galaxies than $L \geq L^*$ galaxies.
Specifically, using only those absorbers with galaxy surveys complete $0.25\,L^*$, we find that {\bf{all}} of \OVI\ absorbers 
studied have physically nearest galaxies in the $0.25L^*<L<L^*$ range. This result suggests 
that the majority of metals expelled into the CGM/IGM originate in sub-$L^*$ galaxies ($0.25L^*<L<L^*$ for the purposes of this study).

% Figure 3

The hypothesis that \OVI\ absorbers are associated primarily with 
low-luminosity galaxies was originally proposed by \citet{tumlinson05} based 
on the $dN/dz$ of \OVI\ absorption systems, a hypothesis that was supported by our
first galaxy-absorber study \citep{stocke06}. Later, \citet{prochaska11} made a 
similar suggestion based on their galaxy survey work, specifically identifying 
sub-$L^*$ galaxies as the primary associated galaxies for low-$z$ \lya\ and 
metal-line systems. More recently, a very deep galaxy survey ($L\geq0.01\,L^*$) 
by \citet{burchett16} found \CIV\ absorption associated primarily with galaxies 
at $L\geq0.3\,L^*$. Although the present galaxy survey is not as deep as that used by \citet{burchett16}, our results are consistent with the intriguing speculation that CGM/IGM metals come primarily from {sub-$L^*$}
galaxies. The combination of the \citet {burchett16} result, which suggests a lower luminosity bound of $L\geq0.3L^*$, and the present study, which suggests an upper bound of $L\leq L^*$, limits the bulk of the metals ejected to a source population in the sub-$L^*$ regime.

Theoretical studies suggest
$L>L^*$ galaxies are too massive to allow metal-enriched gas to easily 
escape beyond \Rvir\, while true dwarfs may eject too little gas 
\textit{in toto} to be major contributors to CGM/IGM ``metal pollution''. 
Since the sub-$L^*$ galaxy population possesses bulk 
metallicities of a few tenths solar values \citep{tremonti04}, which is comparable to 
the absorber metallicities found for photo-ionized CGM absorbers 
\citep{stocke13,werk14,keeney17}, outflows escaping from these modest mass 
galaxies would not require significant dilution by more pristine gas to be 
observed as low-$z$ CGM metal-line systems. 

This study shows that \OVI\ is not spread beyond distances $\sim$ 0.6 Mpc, comparable to or less than the size of a small galaxy group (R$_{vir} \approx$ 1 Mpc for a group halo of 10$^{13.5}$ M$_{\odot}$) 
or width of a large-scale structure filament. Specifically, this distance is significantly greater than the virial radius of even a single 10 L$^*$ galaxy while being comparable to the virial radius of a small group of galaxies with $L\sim 2 L^*$.

The current sample includes absorbers associated with 8 well-studied galaxy groups (5 \OVI\ detections and 3 non-detections) with total luminosities of 4-55L$^*$ \citep[estimated halo masses of 10$^{13.5-15}$ M$_{\odot}$][] {stocke13}. 
While a few \OVI\ detections occur in rather sparse regions (e.g., PKS~0405-123/0.09180) that are in at best very poor groups of galaxies, most of the absorbers in this survey have galaxy densities comparable to the 
8 for which detailed group membership analyses have been done. The absence of metal-bearing absorbers at $\rho >$ 0.6~Mpc, even from quite low luminosity galaxies, argues that galactic winds do not stream freely away from individual
galaxies and groups. For example, even a galactic wind speed of 200 \kms\ will carry metals to $\sim$0.6~Mpc in only 3 billion years if unimpeded by gravity or mass-loading. Other studies \citep {penton04, stocke07} have set only upper limits, of a few percent solar values, on 
metallicities of absorbing gas found several Mpc from the nearest galaxy in ``voids''. This result and the current study support the hypothesis that most or all of the metals produced in galaxies remain within the confines of the galaxy group in 
which the source galaxy is a member. Because more isolated galaxies are poorly represented in this study, we cannot draw any firm conclusions about the spread of metals from those systems.

In conclusion, relatively pristine ($Z<0.1\,Z_{\Sun}$) gas can be accreted by a galaxy group, and the 
metal-enriched gas expelled by the member galaxies does not appear to be transported to distances beyond the group radius. While luminosity function studies of galaxy groups find many more sub-$L^*$ galaxies than $L>L^*$ galaxies (i.e., groups have luminosity functions 
approximately given by Schechter functions), differences in the relative numbers of sub-$L^*$ galaxies in groups are observed \citep{zabludoff00} and could create different metallicity evolution histories between groups if all of their metals are retained inside the group.
This may contribute to the substantial width observed in the mass-metallicity relationship \citep{tremonti04}.

It is likely that both collisionally-ionized (CIE) and photo-ionized (PIE) OVI absorbers are present in this sample. Based on an unbiased \OVI\ sample from \citet{savage14},  $\sim1/3$ of all \OVI\ absorbers at the $N_{\rm O\,VI}$ levels investigated here have inferred $T \geq 10^5$~K. Some of these CIE \OVI\ absorbers likely arise in warm-hot gas which may be associated with entire galaxy groups \citep{stocke14}. However, it is challenging to ascribe any individual absorber unambiguously to either an individual galaxy or to the entire group to which it belongs, since most star-forming galaxies are members of small groups \citep{tully09} and almost all ``passive'' galaxies are members of rich groups or clusters \citep{dressler97}. We have not attempted to make that distinction here but leave this more challenging question to on-going and future investigations which include our own HST study of \HI\- and \OVI\,-absorbing gas associated with rich groups of galaxies \citep[see e.g.,][]{stocke17}.

\acknowledgments
This work was supported by NASA grants NNX08AC146 and NAS5-98043 to the University 
of Colorado at Boulder for the \hst/COS project. CTP, JTS, and BAK gratefully 
acknowledge support from NSF grant AST1109117.

\facilities{AAT (AA$\Omega$), Blanco (HYDRA, MOSAIC), FUSE, HST (COS), MMT (Hectospec), Sloan, WIYN (HYDRA), WIYN:0.9m (MOSAIC)} 
%\software{Numpy}

\clearpage
%\bibliographystyle{yahapj}
%\bibliography{references}

\end{document}